\documentclass[conference]{IEEEtran_Compressed}
\renewcommand\IEEEkeywordsname{Index Terms}

\ifCLASSINFOpdf
\else
  \usepackage[dvips]{graphicx}
  \graphicspath{{./images/}}
  \DeclareGraphicsExtensions{.eps}
\fi
\usepackage[cmex10]{amsmath}
\usepackage{algorithm}
\usepackage{algorithmic}
\usepackage[noadjust]{cite}
\usepackage{makecell}
\usepackage{multirow}

\usepackage{array}
\usepackage{mdwmath}
\usepackage{mdwtab}

\usepackage[caption=false,font=footnotesize]{subfig}

\begin{document}
\title{XJ-BP: Express Journey Belief Propagation Decoding for Polar Codes
\vspace{-10pt}
}

\author{\IEEEauthorblockN{Jingwei Xu, Tiben Che, Gwan Choi}
\IEEEauthorblockA{Department of Electrical and Computer Engineering\\
Texas A\&M University\\
College Station, Texas 77840\\
Email: $\lbrace$xujw07, ctb47321, gchoi$\rbrace$@tamu.edu}}

\maketitle

\begin{abstract}
This paper presents a novel propagation (BP) based decoding algorithm for polar codes.
The proposed algorithm facilitates belief propagation by utilizing the specific constituent codes that exist in the factor graph, which results in an express journey (XJ) for belief information to propagate in each decoding iteration.
In addition, this XJ-BP decoder employs a novel round-trip message passing scheduling method for the increased efficiency.
The proposed method simplifies min-sum (MS) BP decoder by 40.6\%.
Along with the round-trip scheduling, the XJ-BP algorithm reduces the computational complexity of MS BP decoding by 90.4\%;
this enables an energy-efficient hardware implementation of BP decoding in practice.

\end{abstract}

\IEEEpeerreviewmaketitle

\section{Introduction}
Polar codes are proposed by Arikan~\cite{arikan2009channel} as a type of error-correction coding (ECC) method that provably achieves the capacity of symmetric binary-input discrete memoryless channels (B-DMCs).
With its low error-floor performance~\cite{eslami2010bit} and high regularity in coding structure, polar codes attract a significant attention and have the potential to become a standard ECC for the future communication and data storage systems.

There are two widely-considered approaches to decode polar codes.
These are successive cancellation (SC) and belief propagation (BP) algorithms.
The SC algorithm receives more attention because of its low computational complexity $\mathcal{O}(nlogn)$, where $n$ is the code length.
However, decoders based on SC algorithm suffer from the high latency and limited throughput due to their serial decoding natures.
Recently several efforts have been taken into reducing the SC decoding latency \cite{yuan2014low, sarkis2014fast}.
Sarkis et al. utilized the constituent codes that exist in the polar codes to significantly reduce the SC decoding latency by avoiding tree traversals~\cite{sarkis2014fast}.
Although the latency of SC algorithm is substantially improved, the time complexity of it is still $\mathcal{O}(n)$.
Thus with longer polar codes, SC algorithm is still limited in terms of the throughput.
However, polar codes with longer length are more attractive, because the performance of polar codes is superior to other codes at long codeword lengths.

Another approach to decode the polar codes is belief propagation-based (BP) algorithm, which allows decoding in parallel to achieve much higher throughput in dedicated hardware implementation.
Due to its higher computational demand, compared with SC algorithms, BP does not receive much attentions.
The first attempt at implementing BP on field programmable gate array (FPGA) is presented by Pamuk in \cite{pamuk2011fpga}, where the message passing functions are approximated by the min-sum (MS) algorithm for efficient hardware design.
However, the performance of BP decoding is degraded because of the approximations.
Thus, Yuan et al. explored scaled min-sum (SMS) approximation for message passing functions in~\cite{yuan2013architecture} to remedy the performance penalty.
However, compared with MS algorithm, SMS incurs one extra scaling operations in each message passing. 
Yuan et al. further improved the efficiency of SMS BP decoders using early termination in~\cite{yuan2014early}.
On the other hand, by removing unnecessary computations for frozen bits in polar codes, Zhang et al. reduce the complexity for sum-product (SP) BP decoding in~\cite{zhang2014simplified} by around $25\%$ without decoding performance degradation.

This paper presents the XJ-BP decoder that substantially reduces the computational complexity over the conventional BP MS decoding.
Two novel approaches are developed to achieve the improvements.
First approach utilizes specific constituent codes in the factor graph to reduce the decoding complexity.
In this approach, the rules of the belief propagation in each iteration are simplified using the characteristics of the constituent codes.
Secondly, all existing BP decoders schedule the computations in the same manner as mentioned in~\cite{pamuk2011fpga}.
Our approach uses an alternative scheduling method stemming from ideas discussed by Guo et al. at \cite{guo2014enhanced}.
In~\cite{guo2014enhanced}, polar codes are proposed to be concatenated with parity check codes to achieve higher decoding performances.
We describe and compare the two different scheduling methods in this paper to show that our alternative scheduling method is significantly better than the conventionally used one in terms of decoding efficiency.

We show that along with the novel scheduling method, the XJ-BP MS algorithm yields the same decoding performance of the SMS algorithm with $92.8\%$ reduced amount of computations.
Compared with the conventional MS BP decoding, our proposed method does not only reduce the computations by $90.4\%$ but significantly improves the decoding performance.


The rest of this paper is organized as follows: The background of polar codes and its conventional decoding methods are reviewed in the Section~\ref{polar_codes}.
Section~\ref{SBP} describes the proposed algorithm.
Section~\ref{Sched} discusses the two alternative scheduling strategies for BP decoding.
Numerical simulation results of the proposed algorithm and the comparisons with the conventional BP decoding are given in the Section~\ref{results}.
Finally, the paper is concluded in the Section~\ref{conclusions}.

\section{Polar Codes}
\label{polar_codes}

\linespread{1.0}

\subsection{Construction of Polar Codes}

Polar codes are constructed by taking advantage of the polarization effect to achieve the capacity of symmetric channel.
Encoded recursively using the special procedure as discovered in~\cite{arikan2009channel}, the polar codes polarize the post-decoding reliability of the information bits.
An $(n, k)$ polar code is constructed by assigning $k$ information bits and $(n-k)$ '0's at more reliable positions and unreliable positions, respectively.
Those fixed '0' bits are usually referred as frozen bits.
The $n$-bit message bits including frozen bits and information bits are denoted as $\textbf{u}$ in this paper.
The $n$-bit transmitted codeword $\textbf{x}$ is the product of $\textbf{u}$ and the generator matrix $\textbf{G}$, where $\textbf{G}=\textbf{F}^{\oplus m}$.
$\textbf{F}^{\oplus m}$ is the $m$-th Kronecker power of $\textbf{F}= \begin{bmatrix} 1 & 0 \\ 1 &  1 \end{bmatrix}$ and $m=\log_2 n$.

\subsection{Belief propagation decoding}
\label{BP_dec}

Belief propagation decoding is a message passing algorithm that, through the factor graph, refines the estimations of the codeword $\textbf{x}$ or message $\textbf{u}$ in iterations.

\begin{figure}[]
\centering
\subfloat[]{\includegraphics[height=2.4in]{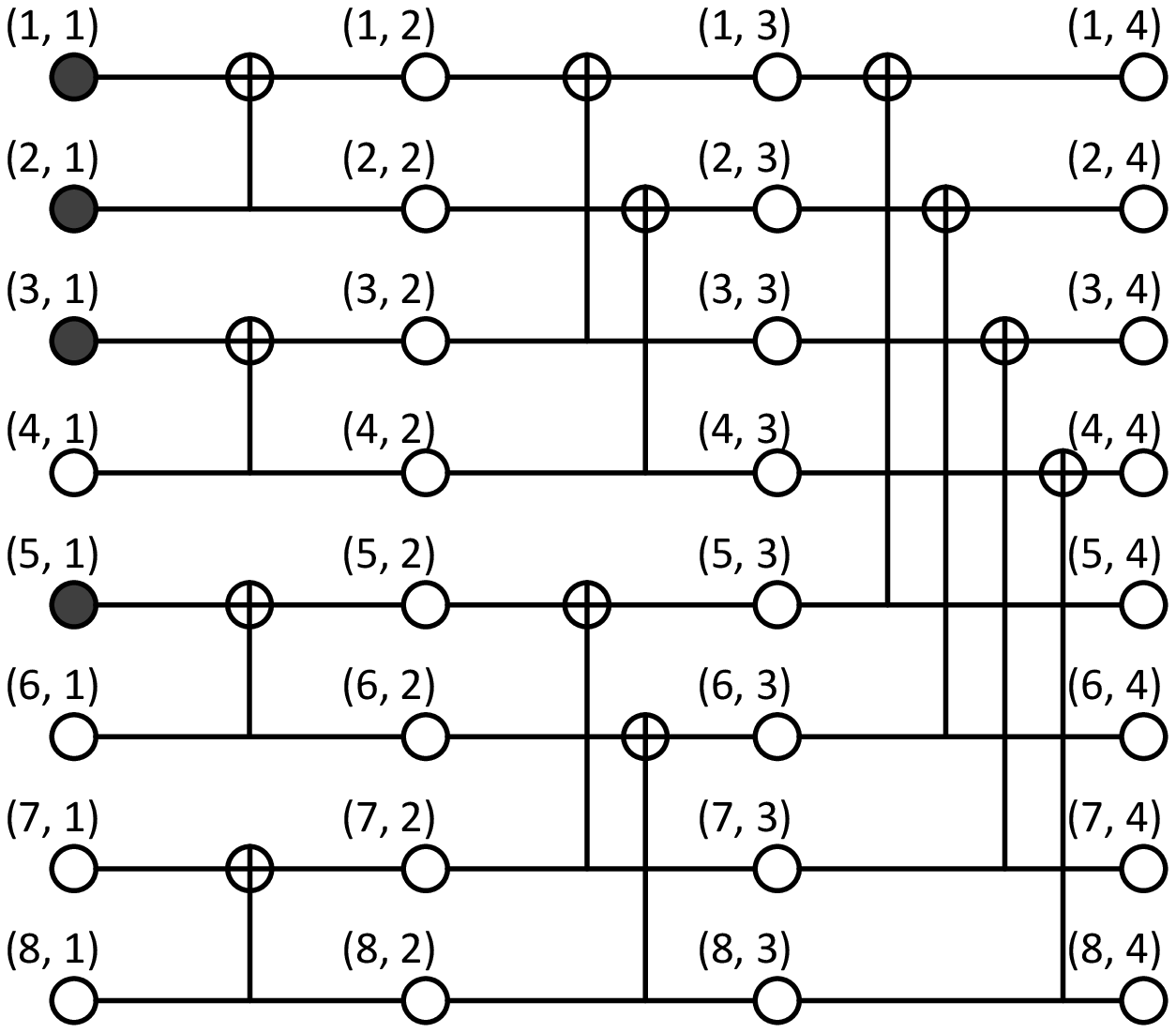}\label{conventional_BP}}
\hfil
\subfloat[]{\includegraphics[width=1.9in]{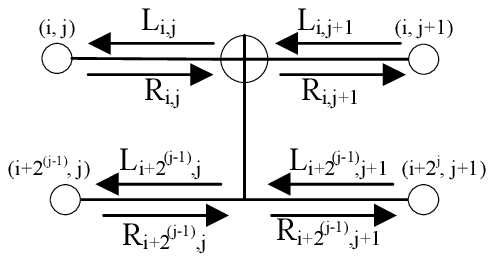}\label{conventional_PE}}
\caption{\protect\subref{conventional_BP} Conventional BP factor graph of $n=8$ polar codes, and \protect\subref{conventional_PE} processing element of conventional BP algorithm.}
\label{system_structures}
\end{figure}

The factor graph of a polar code could be represented by the structure of its encoder.
An example of factor graph of a polar code with $n=8$ is given in Fig.~\ref{conventional_BP}.
As the figure shows, there are $m$ stages in the factor graph, $m=\log_2(n)$.
The bits on the most left column correspond to the message.
In the figure, the black nodes and white nodes in the left column are denoted as the frozen bits and the information bits respectively.
With recursive encoding by the 2-bit polarization unit through the factor graph, the nodes on the most right column correspond to the codeword.
There are two messages passing through each node.
The message propagated from right to left through node $(i,j)$ is designated by $L_{i,j}$.
The other message passed from the other direction is referred as $R_{i,j}$.
Those messages are presented in the log-likelihood ratios (LLRs).
Conventionally, those LLRs are updated through a series of check node processing elements (PE) as shown in~Fig.~\ref{conventional_PE}.
The computations to update LLRs through iterations are written as follows:
\begin{equation}\label{BP_update_eql}
\begin{array}{c}
L_{i,j} = \mathcal{G}(L_{i,j+1},L_{i+2^{j-1},j+1}+R_{i+2^{j-1},j})		\\
L_{i+2^{j-1},j} = \mathcal{G}(R_{i,j}, L_{i,j+1})+ L_{i+2^{j-1},j+1} 
\end{array}
\end{equation}
\begin{equation}\label{BP_update_eqr}
\begin{array}{c}
R_{i,j+1} = \mathcal{G}(R_{i,j},L_{i+2^{j-1},j+1}+R_{i+2^{j-1},j})		\\
R_{i+2^{j-1},j+1} = \mathcal{G}(R_{i,j}, L_{i,j+1})+ R_{i+2^{j-1},j} 
\end{array}
\end{equation}
where $\mathcal{G}(x,y)=\ln{((1+xy)/(x+y))}$ is the propagation function to update messages. 
In practice, the function $\mathcal{G}$ in Eq.~(\ref{BP_update_eql}) and (\ref{BP_update_eqr}) needs to be simplified by  min-sum approximating $\mathcal{G}(x,y)\approx sign(x)sign(y)min(|x|,|y|)$ or scaled min-sum approximating $\mathcal{G}(x,y)\approx \alpha \cdot sign(x)sign(y)min(|x|,|y|)$, where $\alpha$ is the parameter scaling the $\mathcal{G}$ function.

The messages $L_{i,m+1}$ on the most right column are assigned by LLRs from the channel outputs.
The messages $R_{i,1}$ on the first left column are the pre-decoding LLRs of $\hat{\textbf{u}}$.
Decoding starts by assigning $\infty$ and $0$ to the frozen bits and information bits correspondingly.
Those nodes on the most left column are also referred as leaf nodes in this paper.
The BP decoding is performed by operating processing elements from left to right over and over to refine either $L_{i,1}$ or $R_{i,m+1}$ to estimate the transmitted message $\hat{\textbf{u}}$ or transmitted codeword $\hat{\textbf{x}}$ by:
\begin{equation}
\label{llr_u}
	LLR^{\hat{\textbf{u}}}_i = L_{i,1}
\end{equation}
\begin{equation}
\label{llr_x}
	LLR^{\hat{\textbf{x}}}_i = R_{i,m+1} + L_{i,m+1}
\end{equation}
where $LLR^{\hat{\textbf{u}}}_i$ and $LLR^{\hat{\textbf{x}}}_i$ are the log-likelihood ratios of the message $\textbf{u}
$ and the transmitted codeword $\textbf{x}$, respectively.
They are defined as: 
\begin{equation}
\begin{array}{c}
	LLR^{\hat{\textbf{u}}}_i=\ln \frac{P(\textbf{y}|u_i=0)}{P(\textbf{y}|u_i=1)} ,	LLR^{\hat{\textbf{x}}}_i=\ln \frac{P(\textbf{y}|x_i=0)}{P(\textbf{y}|x_i=1)}
\end{array}
\label{def_llr}
\end{equation}
where $P(\textbf{y}|x)$ represents the probability that $\textbf{y}$ is received as $x$ is given in the transmitter.

\subsection{Constituent codes}
As mentioned above, the polar codes are encoded recursively through multiple coding stages.
Thus, any polar code could be regarded as constituted by two shorter polar codes.
For example, in the Fig.~\ref{conventional_BP}, the polar code of bits $\{(i,4)|i=1,2,...,8\}$ comprises the polar code of bits $\{(i,3)|i=1,2,3,4\}$ and the polar code of $\{(i,3)|i=5,6,7,8\}$ with one more stage polarization.
And the polar code of bits $\{(i,3)|i=1,2,3,4\}$ and the polar code of $\{(i,3)|i=5,6,7,8\}$ further consist of shorter polar codes.
Those shorter polar codes which exist in the composition of a polar code are referred as the constituent codes.
Some specific constituent codes are discovered in~\cite{sarkis2014fast} to reduce the latency of SC decoding of polar codes.
In this paper, the exploitation of constituent codes is discussed in simplifying BP decoding algorithms.
The details of the exploration are given in the following.

\section{Simplified belief propagation decoding}

\linespread{1.0}

\label{SBP}
\begin{figure*}[]
\centering
\hspace{-4em}
\subfloat[]{\includegraphics[height=2.35in]{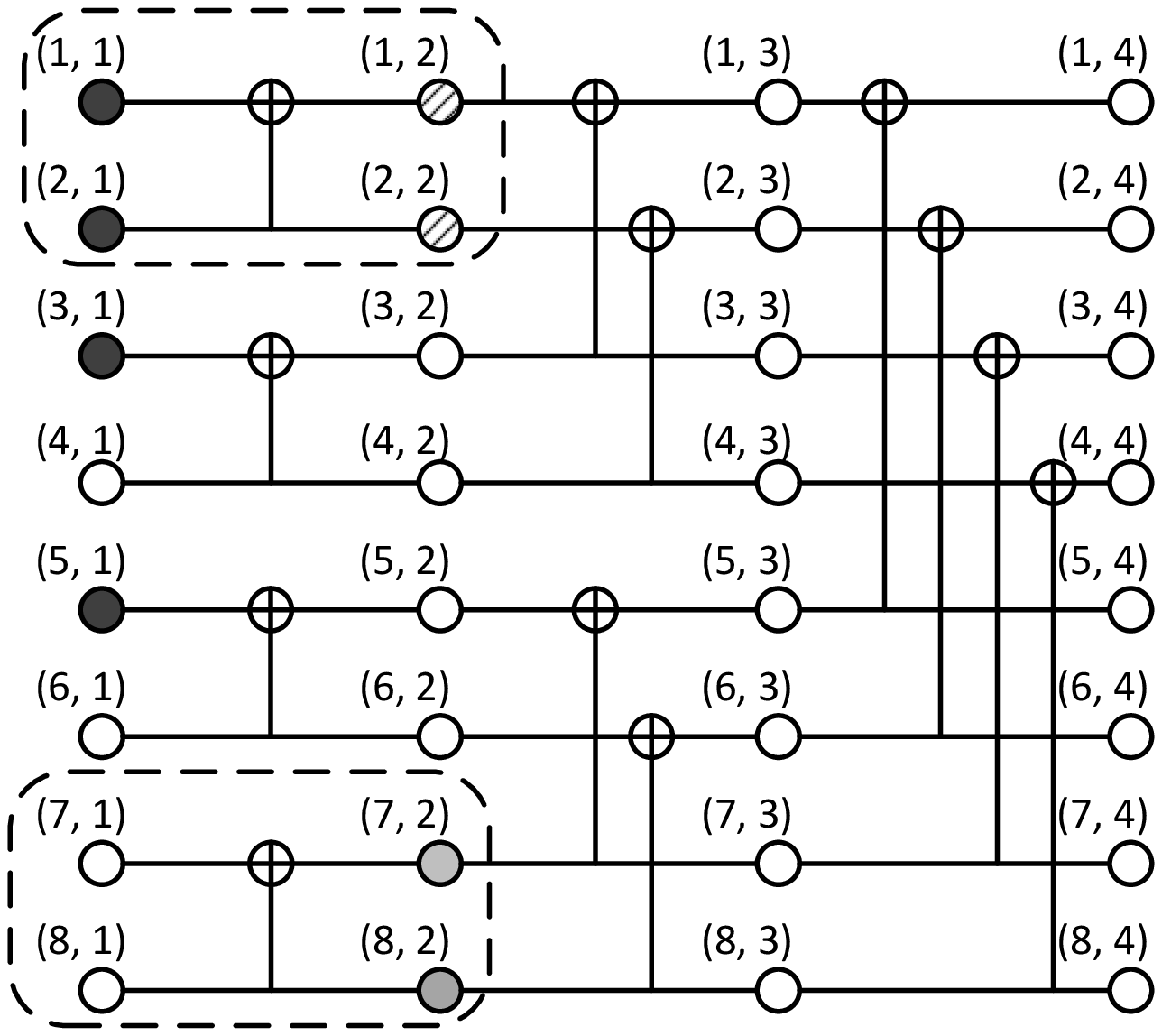}\label{all_frozen_all_info}}
\hspace{1.2em}
\subfloat[]{\includegraphics[height=2.35in]{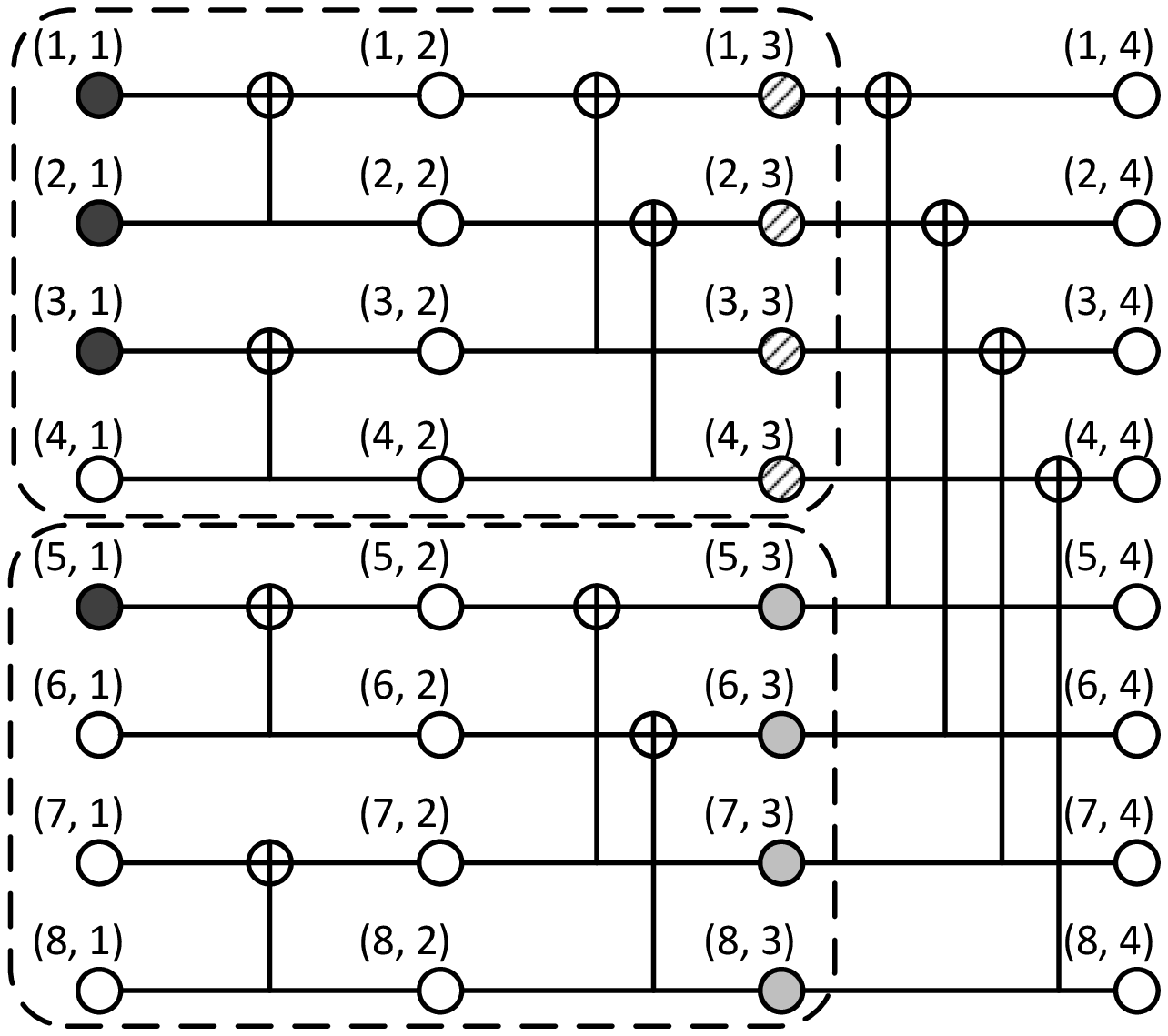}\label{SPC_REP}}
\hspace{0.5em}
\subfloat[]{\includegraphics[height=2.35in]{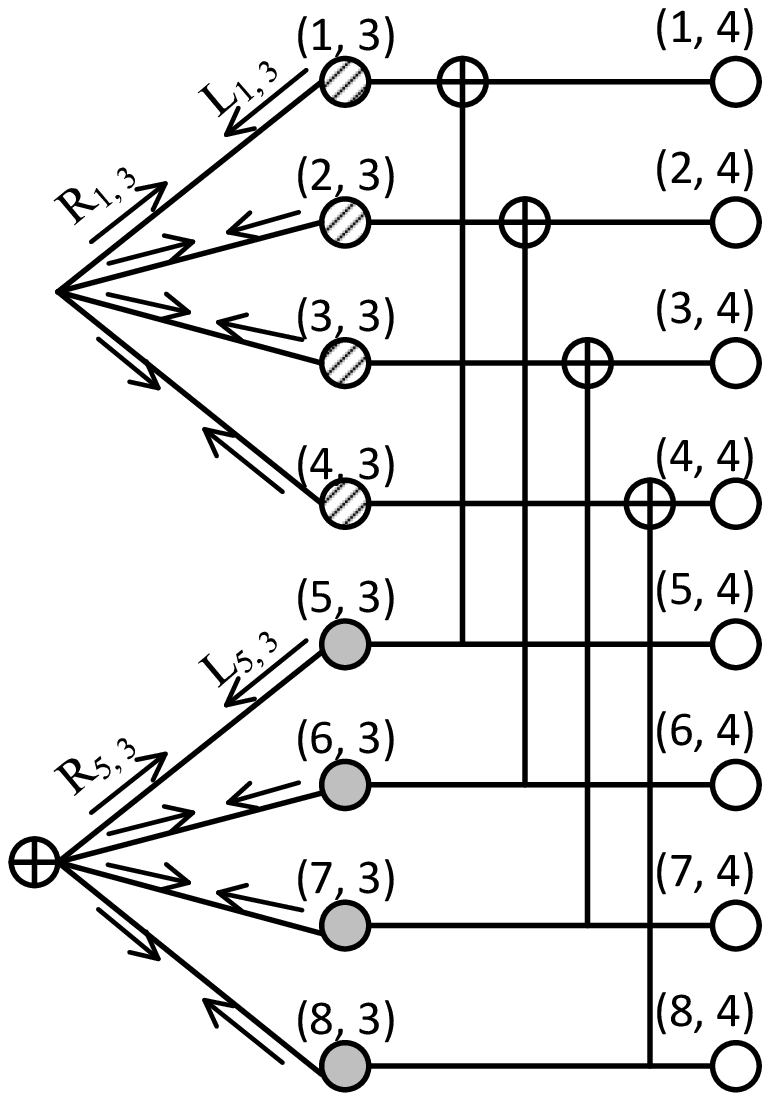}\label{SPC_REP_FG}}
\hspace{-4em}
\caption{\protect\subref{all_frozen_all_info} An example of $\mathcal{N}^{0}$ codes in shadow and $\mathcal{N}^{1}$ codes in gray. \protect\subref{SPC_REP} An example of $\mathcal{N}^{REP}$ codes in shadow and $\mathcal{N}^{SPC}$ codes in gray. And \protect\subref{SPC_REP_FG} the simplified factor graph for the example of $\mathcal{N}^{REP}$ and $\mathcal{N}^{SPC}$ codes.}
\label{ml}
\end{figure*}

In this section, we present different types of constituent codes which can help reduce the complexity of BP decoding algorithm.
The general idea of our algorithm is to refine the estimation of the transmitted codeword $\hat{\textbf{x}}$ without traversing the entire factor graph in each iteration.
The various constituent codes are studied in this section to simplify the factor graph so as to reduce the decoding the complexity.

\subsection{All-frozen $\mathcal{N}^{0}$ codes}
First type of the useful constituent codes are the codes whose left leaf nodes are all frozen bits.
These codes are referred as $\mathcal{N}^{0}$ codes.
Fig.~\ref{all_frozen_all_info} shows an example of $\mathcal{N}^{0}$ code, where the shadowed nodes of $\{(1,2), (2,2)\}$ compose a $\mathcal{N}^{0}$ code.
For those codes, there is no necessity to compute their LLRs, since the codeword is fixed by the frozen bits already.
If the frozen bits are set to $0$, the nodes of $\mathcal{N}^{0}$ codes are also $0$ in the encoding factor graph.
Thus, by setting messages $R_{i,j}$ of nodes $\mathcal{N}^{0}$ codes as $\infty$ before the decoding, the decoding can be performed in each iteration without operating redundant processing elements left to the $\mathcal{N}^{0}$ codes.

\subsection{All-information $\mathcal{N}^{1}$ codes}

As the counterpart of the $\mathcal{N}^{0}$ codes, $\mathcal{N}^{1}$ codes have their all leaf nodes of information bits.
An existence of $\mathcal{N}^{1}$ code in the $n=8$ polar code example is given in Fig.~\ref{all_frozen_all_info}.
In the figure, the grayed codeword $\{(7,2),(8,2)\}$ is a $\mathcal{N}^{1}$ code whose leaf nodes are all information bits.

From the aspect of the factor graph, the refinement does originate from checking information provided by the frozen bits on leaf nodes.
Since there is no frozen bits on the leaf nodes, it is implied that the messages do not get refined by further message passing through $\mathcal{N}^{1}$ codes.
From the Eq.~(\ref{BP_update_eql})~and~(\ref{BP_update_eqr}), it also shows that the $R_{i,j+1}$ and $R_{i+2^{j-1},j+1}$ are not updated with consistent zeros of $R_{i,j}$ and $R_{i+2^{j-1},j}$.
Thus the computations for $\mathcal{N}^{1}$ codes could be removed through BP decoding.


\subsection{Repetition $\mathcal{N}^{REP}$ codes}
Another observation from the factor graph is that there exist considerable amount of constituent codes which only have a single information bit on the last leaf nodes.
Those codes duplicate the only information bit by multiple times to construct the codeword.
The repetition codes are referred as $\mathcal{N}^{REP}$ codes.
The example given in Fig.~\ref{conventional_BP} does contain a $\mathcal{N}^{REP}$ code as shows in Fig.~\ref{SPC_REP}, where the shadowed nodes $\{(1,3),(2,3),(3,3),(4,3)\}$ constitute a $\mathcal{N}^{REP}$ code.

Since we already know that $\mathcal{N}^{REP}$ codes are formed by duplication, the conventional factor graph can be simplified so as to avoid message passing through multiple message stages.
The corresponding example of the factor graph of the $\mathcal{N}^{REP}$ code is given in the Fig.~\ref{SPC_REP_FG}, where the top 4 shadowed nodes constitute a repetition code.
Since each node is a duplication of others, they share the belief messages with others in the factor graph.
The message passing rule of the $\mathcal{N}^{REP}$ codes follows the theory of factor graph \cite{richardson2008modern} as:
\begin{equation}
\label{rep_BP}
	R_{i,j} = \sum\limits_{k \neq i} L_{k,j}
\end{equation}

For a repetition code with length $l$, the complexity of conventional BP is $\mathcal{O}(l\log l)$.
Whereas the complexity of the proposed updating rule is $\mathcal{O}(l)$.
Specifically, the proposed algorithm for length-$l$ $\mathcal{N}^{REP}$ codes takes $(2l-1)$ two-input additions.
Indiscriminately treating nodes of $\mathcal{N}^{REP}$ codes as normal nodes by using conventional BP consumes $(2l\log_2 l)$ comparisons operations and same amount of additions.



\subsection{Single parity check $\mathcal{N}^{SPC}$ codes}
\label{SPC_intro}

The other type of constituent codes exists in polar codes is the single parity check code.
For those constituent codes that only have a single frozen bit on the first leaf node, the codewords are actually single parity check (SPC) codes, the sums of whose codewords are always zero in binary field.
The SPC codes are also referred as $\mathcal{N}^{SPC}$.

As Fig.~\ref{SPC_REP} shows, the leaf nodes of the grayed constituent codeword $\{(5,3),(6,3),(7,3),(8,3)\}$ are all information bits except the first one.
Similar to $\mathcal{N}^{REP}$ codes, it is unnecessary to evaluate through all conventional computations to update the messages $R$ of those nodes.
Since the codeword is a SPC code, the factor graph of the $\mathcal{N}^{SPC}$ codes could be modeled as a parity check node connected with all bits of the codeword.
The modified factor graph of the $\mathcal{N}^{SPC}$ code in the example is shown in Fig.~\ref{SPC_REP_FG}.
In the figure, an additional parity check nodes is added to propagate the belief information among the nodes.
With the consistency on using min-sum algorithm, the parity check update is written as:
\begin{equation}
\label{spc_BP}
	R_{i,j} = \prod\limits_{k \neq i} sgn(L_{k,j}) \cdot \min\limits_{k \neq i} \vert L_{k,j} \vert
\end{equation}

Similar as the repetition codes, the complexity of the modified message passing algorithm is $\mathcal{O}(l)$ for length-$l$ single parity check code which is superior to the complexity of the conventional algorithm, $\mathcal{O}(l \log l)$.
Thus with longer constituent codes, more computation are saved with the proposed algorithm.

Noticeably, the $\mathcal{N}^{0}$ and $\mathcal{N}^{1}$ codes are not usually included in $\mathcal{N}^{SPC}$ and $\mathcal{N}^{REP}$ codes in reality.
Simplifications of message passing on those four different types of constituent codes are all applied simultaneously.
The distributions of exclusive constituent codes in a $(1024, 512)$ are shown in Table~\ref{codes_sizes}.
As the table shows, there are considerable amount of constituent codes in the polar code.
There are more number of $\mathcal{N}^{REP}$ and $\mathcal{N}^{SPC}$ codes than $\mathcal{N}^{0}$ and $\mathcal{N}^{1}$ codes.
Thus an efficient BP algorithm design for the $\mathcal{N}^{REP}$ and $\mathcal{N}^{SPC}$ codes could substantially further reduce the BP decoding complexity.
Also notice that the distribution of the constituent codes does also depend on the code rate and polar codes with rate of $0.5$ contain relatively less number of constituent codes.
With higher code rate, it is more attractive to apply the proposed methods to simplify the message passing.
The details of complexity analysis will be presented in Section~\ref{Comp_Analysis}.

\begin{table}[h]
\setlength{\extrarowheight}{2pt}
\small
\centering
\caption{Number of all constituent codes with different sizes in a (1024, 512) polar code with rate of $0.5$}
\begin{tabular}{cccccccc}
\Xhline{1.2pt}
\multirow{2}{*}{} & \multicolumn{6}{c}{Constituent codes sizes} & \multirow{2}{*}{All} \\ \cline{2-7}
                  & 4     & 8    & 16   & 32   & 64   & 128   &                      \\ 
\Xhline{1.2pt}
$\mathcal{N}^{0}$                & 3     & 3    & 2    & 2    & 0    & 1     & 11                   \\
$\mathcal{N}^{1}$                 & 3     & 3    & 2    & 1    & 0    & 0     & 9                    \\
$\mathcal{N}^{REP}$                 & 16    & 8    & 4    & 1    & 1    & 1     & 31                   \\
$\mathcal{N}^{SPC}$                 & 15    & 5    & 3    & 1    & 1    & 0     & 25                   \\ 
\Xhline{1.2pt}
\end{tabular}
\label{codes_sizes}
\end{table}

With the constituent codes applied to reduce computations, the journey for message passing is simplified so that the LLRs of $\hat{\textbf{u}}$ are not immediately available from BP iterations.
Thus in the proposed algorithm, we focus on refining the estimations of transmitted codeword $\hat{\textbf{x}}$ instead of messages $\hat{\textbf{u}}$.
The estimated LLRs of $\hat{\textbf{x}}$, the soft estimations of transmitted codeword $\textbf{x}$ in log likelihood ratio, are represented by Eq.~(\ref{llr_x}).
As aforementioned, $L_{i,m+1}$ are LLRs from the channel outputs.
So in our algorithm, $R_{i,m+1}$ is refined in iterations to accomplish decoding.
The details how the computations are scheduled to accommodate the simplification is presented in the next section.

\section{Scheduling}
\label{Sched}
This section presents the two different ways to schedule the computations of conventional BP decoding algorithm.
Next the scheduling plan for the proposed BP decoding is illustrated.
Finally, we present a method to terminate early the BP decoding iterations.

\subsection{Round-trip BP updating}
\begin{figure}[]
\centering
\subfloat[]{\includegraphics[width=3.5in]{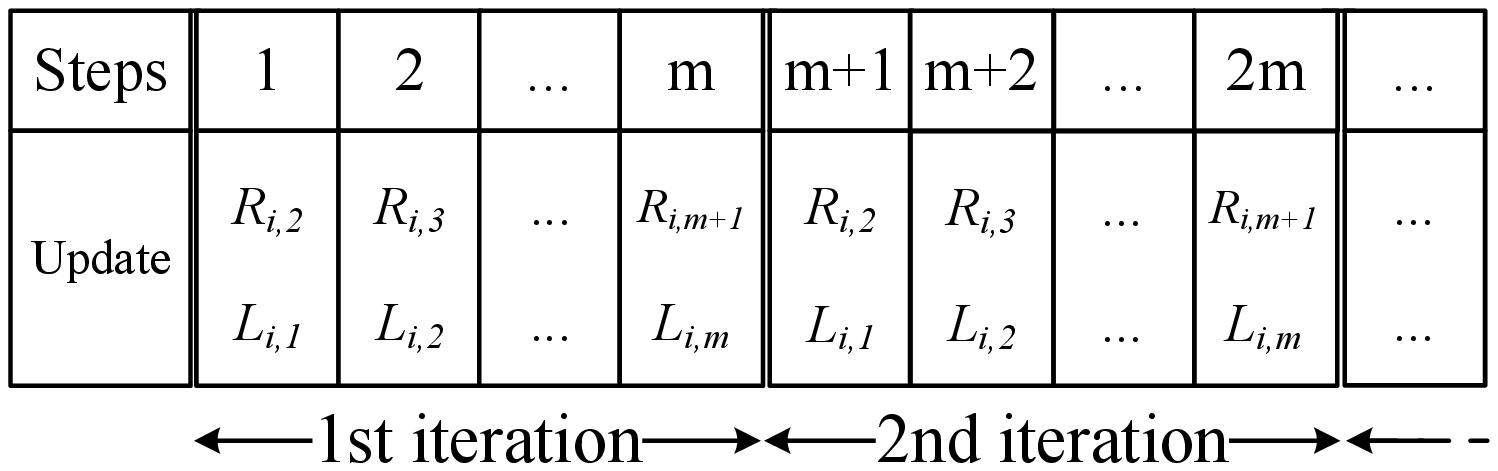}\label{tradsche}}
\hfil
\subfloat[]{\includegraphics[width=3.5in]{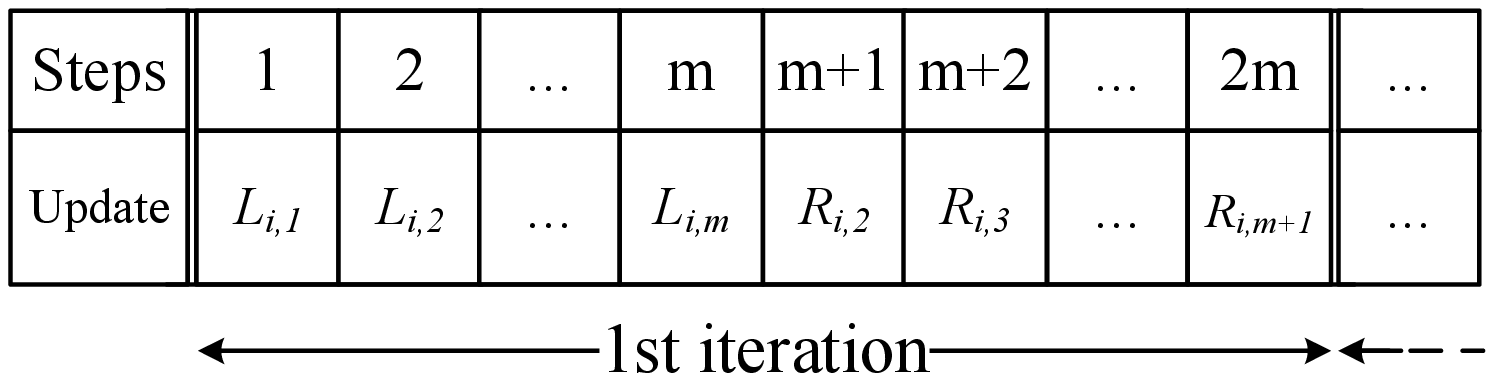}\label{roundtrip}}
\caption{Two types of scheduling methods in BP decoders. \protect\subref{tradsche} Computations scheduled in the conventional BP decoders, and \protect\subref{roundtrip} Computations scheduled in a round-trip updating fashion.}
\label{scheduling}
\end{figure}

The computations of all existing conventional BP decoders are based on the processing element of Fig.~\ref{conventional_PE}.
In the other proposed BP processing elements, the messages are computed simultaneously for both directions of left-to-right and right-to-left.
Fig.~\ref{tradsche} shows the computations scheduled by the conventional BP decoding.
As the figure shows, each iteration consists of $m$ stages of computations, where $m=\log_2(n)$ is the number of stages in the factor graph.
For each stage, the messages of both direction $R_{i+1,j}$ and $L_{i,j}$ of each stage are computed.
And the computations are repeated in one-way direction from left to right iteratively.
However, this scheduling method lacks efficiency.
For instance, it is inefficient to update $L_{i,1}$ in step 1 before having updated $L_{i,2}$ in step 2.

Another way to schedule the computations is to separately update right-to-left messages and left-to-right messages.
Fig.~\ref{roundtrip} shows the schedule of messages updated in this fashion.
As the figure shows, the computations of each iteration are separated to two parts.
In the first part, the $L_{i,j}$ messages are updated from column $m+1$ to the most left nodes existing in the modified factor graph.
The second is following to update the other direction message $R_{i,j}$ from left to the column $m+1$.
Since in each iteration there is a round trip through the factor graph, this scheduling scheme is referred as round-trip scheduling in this paper.
Though each iteration of this modified scheduling contains a round-trip visit of nodes instead of one-way traverse, the amount of computations is same as that of the conventional scheduling, because only half of messages, either $L_{i,j}$ or $R_{i,j}$, are updated in each direction.
Furthermore, the round-trip scheduling significantly improves the efficiency in terms of number of iterations.
Section~\ref{Comp_Analysis} will discuss the number of iterations in details.

In this paper, we employ the proposed round-trip scheduling to update $R_{i,m+1}$ as discussed above in order to promote the efficiency.
In contrast with conventional BP decoding, for constituent $\mathcal{N}^{REP}$ and $\mathcal{N}^{SPC}$ codes, Eq.~(\ref{rep_BP}) and (\ref{spc_BP}) instead of Eq.~(\ref{BP_update_eqr}) are used to update messages $R_{i,j}$.

\subsection{Early Termination}
\label{ET}
In this paper, we apply early termination technique to determine whether the decoding is successfully done or not.
Polar codes belong to the block codes.
For block codes, $H$ matrix could be used for codeword detection.
According to the coding theory~\cite{moon2005error}, the parity check matrix $H$ could be derived given generator matrix $G^{\prime}$.
Here $G^{\prime}$ is a $k \times n $ matrix consisting rows of matrix $\textbf{G}$ corresponding to the positions of the information bits.
Then the termination of a decoding is indicated by the equation:
\begin{equation}
\label{EarlyTermination}
	\hat{\textbf{x}} H = \textbf{0}
\end{equation}
where $\hat{\textbf{x}}$ is the hard decision of the transmitted codeword estimations, i.e.
\begin{equation}
\hat{\textbf{x}}_i=
\begin{cases}
0,\quad	LLR^{\hat{\textbf{x}}}_i > 0\\
1,\quad	otherwise
\end{cases}
\end{equation}

\section{Simulation and Discussion}
\linespread{1.0}
\label{results}
In this section, we set up simulations to verify the proposed algorithm.
Compared with the conventional BP decoding algorithm, the complexity and performance of the proposed algorithm are also analyzed and discussed in this section.
As an example, $(1024, 512)$ polar code is used to emulate the proposed decoder with max number of iterations of 60.

\subsection{Decoding Performance}
\label{DecodingPerf}
\begin{figure}[!t]
\centering
\includegraphics[width=3.75in]{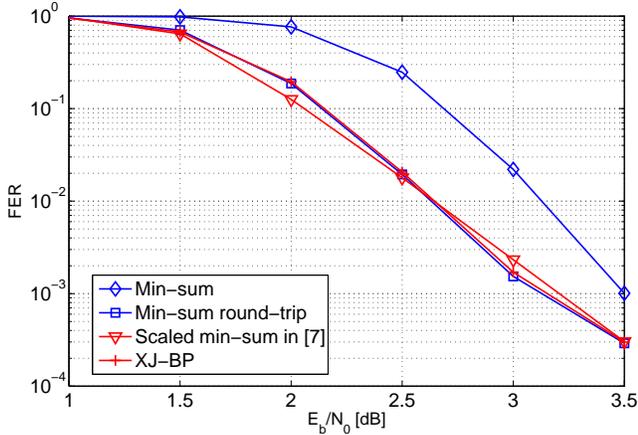}
\caption{Decoding performance of the proposed BP decoding algorithm for $(1024, 512)$ polar code with $\text{rate}=0.5$ and max number of iteration of $60$.}
\label{Decodefig}
\end{figure}

Fig.~\ref{Decodefig} shows the decoding performances of four decoding strategies.
They are the conventional min-sum (MS) BP algorithm with conventional scheduling, the conventional MS BP algorithm with round-trip scheduling, the scaled min-sum (SMS) algorithm proposed in~\cite{yuan2014early} with conventional scheduling and the proposed algorithm.
As the results show, the min-sum BP decoding with the round-trip computation scheduling considerably outperforms the conventional min-sum algorithm.
The performance of the min-sum BP algorithm with round-trip updating is very close to that of the scaled min-sum algorithm~\cite{yuan2014early}.
%

We also show that the proposed XJ-BP algorithm yields almost same performance as the conventional BP algorithm with round-trip scheduling does.
It means that the simplifications for constituent codes do not result in any degradation in decoding performance.

\subsection{Computation Complexity Analysis}
\label{Comp_Analysis}
After showing the decoding performance of the proposed algorithm, here we discuss the complexity reduction by the proposed XJ-BP algorithm.

\begin{figure}[!t]
\centering
\includegraphics[width=3.75in]{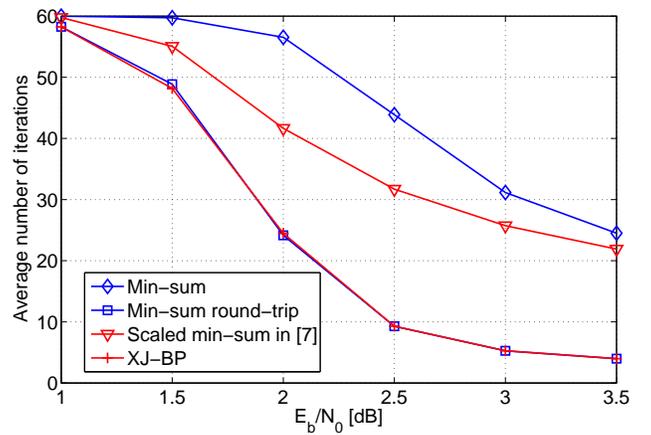}
\caption{Average numbers of iterations of the proposed BP decoding algorithm for $(1024, 512)$ polar code with $\text{rate}=0.5$.}
\label{Iter}
\end{figure}

First of all, the average numbers of iterations of those algorithms are summarized in the Fig.~\ref{Iter}.
It is shown in the figure that with the round-trip scheduling computations, the efficiency of the BP algorithm is significantly increased.
Noticeably scaled min-sum BP algorithm reduces the number of iterations.
However the reduction is at the cost of the additional scaling computation in each node update.
The interesting phenomenon from this experiment is that the round-trip scheduling significantly improves the iteration efficiency without the additional computational complexity cost.
Under the condition of high $E_b/N_0 = 3.5$, the round-trip BP scheduling only takes  $3.98$ average iterations to complete decoding.
As mentioned in Section~\ref{Sched}, the amounts of computations for conventional scheduling and round-trip scheduling in each iteration are the same.
Compared with $24.5$ average number of iterations consumed by the conventional MS BP decoding, the decoding efficiency is immediately improved by $83.7\%$ without considering the simplification on factor graph yet.
Also, it is addressed that the proposed XJ-BP algorithm does not reduce the number of iterations compared with the traditional BP but with round-trip scheduling.

Secondly, we evaluate the reduction of computations in each iteration resulting from the proposed XJ approach for message passing.
As mentioned above, computations for nodes of $\mathcal{N}^{0}$ and $\mathcal{N}^{1}$ codes could be removed directly.
The computations of $\mathcal{N}^{REP}$ and $\mathcal{N}^{SPC}$ codes are reduced by XJ-BP.
The numbers of total operations (2-input addition or 2-input comparison) are shown in the Table.~\ref{comp}.
In the table, polar codes are set at $\text{rate}=0.5$ and the channel polarization is done under the binary erasure channel (BEC) model with erasure ratio of $0.3$.
It is shown that the total number of computations could be reduced by about 40\% in each iteration using the proposed simplified BP algorithm.
And we found that this ratio is kept at about 40\% even with significantly longer code length.
In another word, the proposed simplification saves around $40\%$ amount of computations regardless of lengths of the polar codes.

\begin{table}[t]
\setlength{\extrarowheight}{2pt}
\small
\centering
\caption{Number of computations of XJ-BP algorithm with all polar codes at $\text{rate}=0.5$}
\begin{tabular}{cccccc}
\Xhline{1.2pt}
\multirow{2}{*}{}                  & \multicolumn{5}{c}{Polar code sizes}      \\ \cline{2-6} 
                                   & 128    & 256    & 512    & 1024   & 2048  \\ \Xhline{1.2pt}
Conventional BP                    & 1792   & 4096   & 9216   & 20480  & 45056 \\
XJ-BP                        & 1040   & 2488   & 5536   & 12160  & 27304 \\
Ratios [\%] & 58.0\% & 60.9\% & 60.1\% & 59.4\% &  60.6\%      \\ \Xhline{1.2pt}
\end{tabular}
\label{comp}
\end{table}

Another factor that affects the simplification is the code rate.
Table.~\ref{rate} shows the number of computations for proposed algorithm decoding a polar code of length 1024 at different typical code rates.
As the table shows, the proposed algorithm saves more computation resource to decode polar code with higher code rates.
This is because that more constituent codes exist in the factor graph with more unbalanced number of frozen bits and information bits.

\begin{table}[!t]
\setlength{\extrarowheight}{2pt}
\small
\centering
\caption{Computations of XJ-BP algorithm in each iteration at different code rates}
\begin{tabular}{cccccc}
\Xhline{1.2pt}
\multirow{2}{*}{}  & \multicolumn{5}{c}{Code Rates}             \\ \cline{2-6} 
                   & 1/2    & 2/3    & 3/4    & 5/6    & 7/8    \\ \Xhline{1.2pt}
conventional BP    & 20480  & 20480  & 20480  & 20480  & 20480  \\
XJ-BP        & 12160  & 11488  & 10680  & 9376   & 8936   \\
Ratios {[}\%{]} & 59.4\% & 56.1\% & 52.3\% & 45.8\% & 44.6\% \\ \Xhline{1.2pt}
\end{tabular}
\label{rate}
\end{table}

\begin{figure}[!t]
\centering
\includegraphics[width=3.75in]{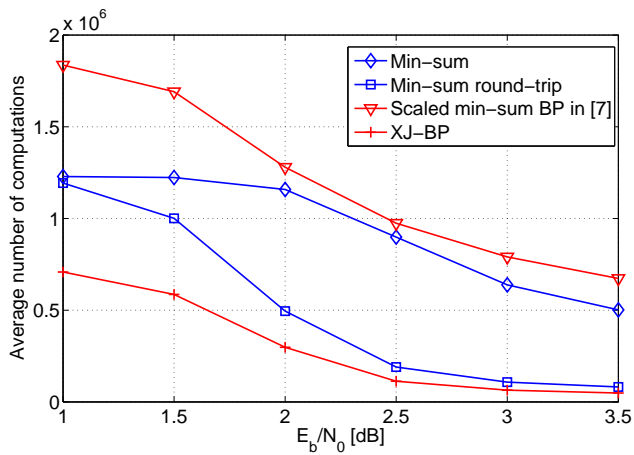}
\caption{Average numbers of computations consumed to decode each codeword of by the proposed BP decoding algorithm for $(1024, 512)$ polar code with $\text{rate}=0.5$.}
\label{overall_comp}
\end{figure}

Finally, the overall complexity reduction is evaluated by considering both the reduced number of iterations and simplified computations in each iteration.
Take the (1024, 512) codes as an example, Fig.~\ref{overall_comp} shows the average numbers of computations to decode one codeword at different levels of $E_b/N_0$.
Due to the extra scaling operations, SMS consumes around $34\%$ more computations over the conventional MS decoding algorithm, although SMS outperforms conventional BP in terms of decoding performance.
Compared with conventional BP decoding, the round-trip scheduling reduces the number of computations by $83.7\%$ at $E_b/N_0=3.5$ resulting from the reduced number of iterations.
Based on round-trip scheduling, the proposed method does not yield any further improvement on number of necessary iterations.
However the XJ-BP decoding simplifies factor graph so as to reduce the computations in each iteration by $40.6\%$.
As a results, the overall complexity is reduced by $90.4\%$ using XJ-BP, compared with conventional BP decoding.

%

\subsection{Discussions}
%
%

From the aspect of practical implementation, the conventional BP processing element symmetrically computes updates for messages $R_{i,j}$ and $L_{i,j}$.
Traditional computations for $R_{i,j}$ as shown in Eq.~(\ref{BP_update_eqr}) are as same as those for $L_{i,j}$ in Eq.~(\ref{BP_update_eql}).
In practical implementation for the proposed algorithm, the processing elements should be designed as only to deal with functions $\mathcal{G}(x, y+z)$ and $\mathcal{G}(x, y)+z$ to satisfy only one-direction message computations.

The message updating rules are different between normal nodes and nodes of the constituent codes in mathematics.
But the basic operations of additions and comparisons for them are similar.
Thus the proposed processing elements could be multiplexed between normal and specific constituent codes.

\section{Conclusion}
\label{conclusions}
In this paper, a novel method is proposed to simplify belief propagation decoding algorithms for polar codes.
By modifying the BP rules for the specific constituent codes, the proposed method significantly simplifies the factor graph of message passing in each iteration.
Additionally, a novel round-trip scheduling approach is developed based on the observations that BP decoding algorithm works more efficiently with it.
The computational efficiencies of different BP-based decoding strategies are evaluated by counting numbers of basic operations.
The results show that the proposed XJ-BP algorithm reduces the computational complexity of MS BP decoding by 90.4\% while yielding the same performance as that of the SMS BP decoding algorithm.

\linespread{0.93}

\bibliographystyle{IEEEtran}
\bibliography{IEEEabrv}

\end{document}